\documentclass[11pt,a4paper]{article}
\usepackage{amsmath,amssymb}
\usepackage{epsfig,graphicx}
\usepackage{subfigure}
\usepackage{graphicx}
\usepackage{rotating}
\usepackage{bm}
\usepackage{axodraw}
\usepackage{color}
\usepackage[sort&compress,square]{natbib}

\newcommand{\mg}{m_{\gamma^{\prime}}}

\begin{document}
\date{\mbox{ }}
\title{{\normalsize  IPPP/07/38; DCPT/07/76;  DESY 07-099\hfill\mbox{}\hfill\mbox{}}\\
\vspace{2.5cm} \Large{\textbf{A Cavity Experiment to Search for Hidden Sector Photons}}}
\author{Joerg Jaeckel$^1$\footnote{{\bf
e-mail}: joerg.jaeckel@durham.ac.uk} \  and Andreas Ringwald$^2$\footnote{{\bf
e-mail}: andreas.ringwald@desy.de}
\\[2ex]
\small{\em $^1$Centre for Particle Theory, Durham University, Durham DH1 3LE, United Kingdom}\\[1.5ex]
\small{\em $^2$Deutsches Elektronen Synchrotron, Notkestra\ss e 85, 22607 Hamburg, Germany} }
\date{}
\maketitle

\begin{abstract}
\noindent We propose a cavity experiment to search for low mass
extra U(1) gauge bosons with gauge-kinetic mixing with the ordinary
photon, so-called paraphotons. The setup consists of two microwave
cavities shielded from each other. In one cavity, paraphotons are
produced via photon-paraphoton oscillations. The second, resonant,
cavity is then driven by the paraphotons that permeate the shielding
and reconvert into photons. This setup resembles the classic ``light
shining through a wall'' setup. However, the high quality factors
achievable for microwave cavities and the good sensitivity of
microwave detectors allow for a projected sensitivity for
photon-paraphoton mixing of the order of $\chi\sim 10^{-12}$ to
$10^{-8}$, for paraphotons with masses in the $\mu{\rm{eV}}$ to
${\rm{meV}}$ range  -- exceeding the current laboratory- and
astrophysics-based limits by several orders of magnitude. Therefore,
this experiment bears significant  discovery potential for hidden
sector physics.
\end{abstract}

\vspace{3ex}

Extensions of the standard model often contain extra U(1) gauge degrees of freedom.
If the corresponding additional gauge bosons have direct renormalizable couplings to
standard model matter,
they are usually referred to as $Z^{\prime}$-bosons. Negative collider searches for the latter have constrained
their mass to   $m_{Z^\prime}\gtrsim {\rm{few}}\times 100\, {\rm{GeV}}$, for couplings of
weak or electromagnetic strength~\cite{Yao:2006px}.

However, in many cases, notably in realistic string-based scenarios
standard model matter is uncharged under the
additional U(1) symmetry and the corresponding gauge boson belongs
to a ``hidden sector'', typically interacting with the standard
model particles only via feeble gravity-like interactions. In these
cases,  the only renormalizable interaction with the standard model
visible sector can occur via mixing~\cite{Okun:1982xi,Holdom:1985ag}
of the photon $\gamma$ with the hidden sector photon
$\gamma^\prime$, often dubbed ``paraphoton''. Clearly, the
sensitivity of collider experiments to photon mixing  is extremely
limited, in particular if the hidden sector photon has a small mass
in the sub-eV range. Presently, the best laboratory limits on a low
mass paraphoton and its mixing with the photon arise from
Cavendish-type tests of the Coulomb
law~\cite{Williams:1971ms,Bartlett:1988yy} and from the search for
signals of $\gamma$--$\gamma^\prime$ oscillations with  laser
``light shining through a wall''
experiments~\cite{Cameron:1993mr,Ahlers:2007rd}. The best
astrophysical limits come from considerations of the energy balance
of stars, in particular the sun, and  from the
non-observation of photon regeneration in
helioscopes~\cite{Popov:1991,Popov:1999}.

In this letter, we propose a laboratory experiment to search for signatures of $\gamma$--$\gamma^\prime$
oscillations by exploiting high-quality microwave cavities. Our setup seems to be realizable with current
technology and has a large window of opportunity for the discovery of low mass,
$\mu{\rm eV}\lesssim m_{\gamma^\prime}\lesssim {\rm meV}$,
hidden sector photons, exceeding the current limits on $\gamma$--$\gamma^\prime$ mixing by several orders
of magnitude.

For definiteness, we will consider an extension of the standard model where one has, at low energies,
say much below the electron mass,
in addition to the familiar electromagnetic {U(1)$_{_\mathrm{QED}}$},  another
hidden-sector {U(1)$_\mathrm{h}$} under which all standard model particles have zero charge.
This may occur quite generally in string embeddings of the standard model
(for general reviews, see e.g.
Refs.~\cite{Quevedo:2002fc,Abel:2004rp,Blumenhagen:2006ci,Marchesano:2007de}),
no matter whether they are based
on compactifications of heterotic (e.g.~\cite{Buchmuller:2006ik,Kim:2007mt}), IIA
(e.g.~\cite{Blumenhagen:2000wh}), and IIB (e.g.~\cite{Aldazabal:2000sa}) string theory. The most general
renormalizable Lagrangian describing these two U(1)'s at low energies is
\begin{equation}
\label{lagrangian}
{\mathcal{L}}= -\frac{1}{4} F^{\mu\nu}F_{\mu\nu}-\frac{1}{4}B^{\mu\nu}B_{\mu\nu}
-\frac{1}{2}\chi\,F^{\mu\nu}B_{\mu\nu}  +\frac{1}{2}m_{\gamma^\prime}^2 B_\mu B^\mu,
\end{equation}
where $F_{\mu\nu}$ is the field strength tensor for the ordinary
electromagnetic {U(1)$_{_\mathrm{QED}}$} gauge field $A^{\mu}$, and
$B^{\mu\nu}$ is the field strength for the hidden-sector
{U(1)$_\mathrm{h}$} field $B^{\mu}$, i.e., the paraphoton.  The
first two terms are the standard kinetic terms for the photon and
paraphoton fields, respectively. Because the field strength itself
is gauge invariant for U(1) gauge fields, the third term is also
allowed by gauge and Lorentz symmetry.  This term corresponds to a
non-diagonal kinetic term, a so-called kinetic
mixing~\cite{Holdom:1985ag}. From the viewpoint of a
low-energy effective Lagrangian, $\chi$ is a completely arbitrary
parameter. Embedding the model into a more fundamental theory, it is
plausible that $\chi=0$ holds at a high-energy scale related to the
fundamental theory. However, integrating out the heavy quantum
fluctuations generally tends to generate non-vanishing~$\chi$ at low
scales. Indeed, kinetic mixing arises quite generally both in field
theoretic~\cite{Holdom:1985ag} as well as in string
theoretic~\cite{Dienes:1996zr,Lust:2003ky,Abel:2003ue,Batell:2005wa,Blumenhagen:2006ux,Abel:2006qt}
setups.
The last term in the Lagrangian~(\ref{lagrangian}) accounts for a possible mass of
the paraphoton. This may arise from the breaking of the paraphoton {U(1)$_\mathrm{h}$} via a Higgs
mechanism and choosing unitary gauge, or, alternatively, may be just an explicit
St\"uckelberg mass term~\cite{Stueckelberg:1938}.

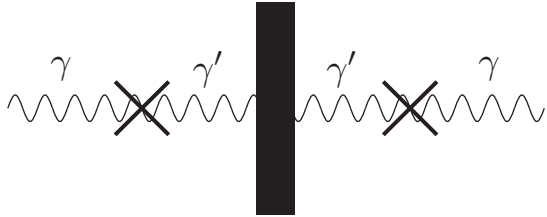
\begin{figure}[t]
\begin{center}
\begin{picture}(190,105)(0,0)
\Photon(0,60)(100,60){5}{9}
\Photon(100,60)(200,60){5}{9}
\linethickness{0.5cm}
\put(100,20){\line(0,1){80}}
\SetWidth{1.5}
\Line(40,50)(60,70)
\Line(40,70)(60,50)
\Line(140,50)(160,70)
\Line(140,70)(160,50)
\Text(20,75)[c]{\scalebox{1.275}[1.275]{$\gamma$}}
\Text(75,75)[c]{\scalebox{1.275}[1.275]{$\gamma'$}}
\Text(125,75)[c]{\scalebox{1.275}[1.275]{$\gamma'$}}
\Text(180,75)[c]{\scalebox{1.275}[1.275]{$\gamma$}}
\end{picture}
\end{center}
\vspace{-1.0cm} \caption{\small Schematic picture of a
``light shining through a wall'' experiment.
The crosses denote the non-diagonal mass terms that convert
photons into paraphotons. The photon $\gamma$ oscillates into the
paraphoton $\gamma'$ and, after the wall, back into the photon
$\gamma$ which can then be detected.}
\label{lswnomagnetic}
\end{figure}

\begin{figure}[t]
\begin{center}
\includegraphics*[bbllx=26,bblly=222,bburx=583,bbury=610,width=.7\textwidth]{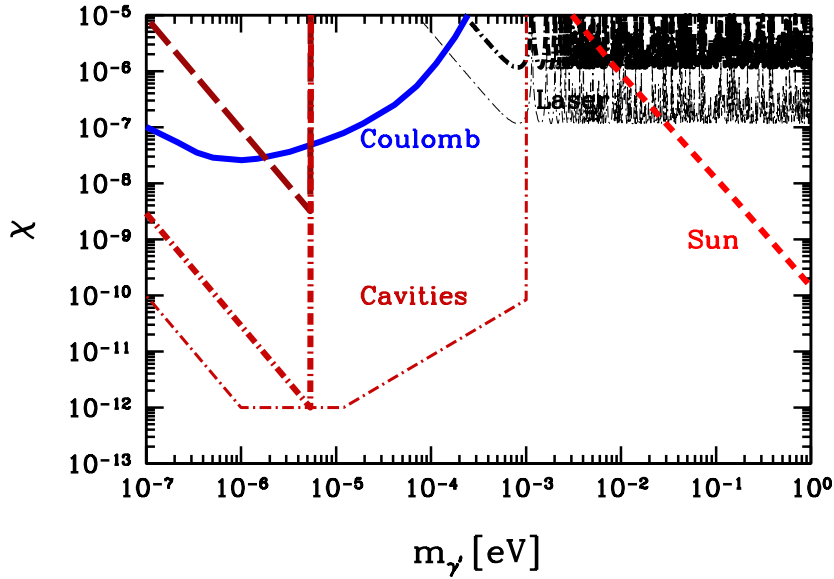}
\caption[...]{Existing bounds on the existence of massive
paraphotons with kinetic mixing and projected sensitivity for the
proposed experiment. Upper limit on the mixing parameter $\chi$
versus the mass $m_{\gamma^\prime}$, obtained from the
non-observation of deviations from Coulomb's
law~\cite{Williams:1971ms,Bartlett:1988yy} (blue, labelled
``Coulomb''), from the non-observation of laser ``light shining
through a wall'' (black, labeled ``Laser''; thick: published limit
from BFRT~\cite{Cameron:1993mr}; thin: projected sensitivity of
ongoing experiments~\cite{Ahlers:2007rd}), and from solar energy
balance considerations~\cite{Popov:1991,Popov:1999} (red, labelled
``Sun''). Also shown is the projected sensitivity of our proposed
``microwaves permeating through a shielding'' setup (darkred,
labelled ``Cavities''). The dashed dotted line corresponds to the
optimistic scenario ($Q=Q^\prime =10^{11}$, $\mathcal{P}_{\rm
em}\sim 1$~W, $\mathcal P_{\rm detectable}=10^{-26}$~W,
$\nu_0=1.3\,{\rm{GHz}}$, i.e. $\omega_0\approx 5.4\, \mu {\rm{eV}}$)
and the dashed fat line to the more modest one ($Q=10^{10}$,
$Q^\prime =10^{4}$, $\mathcal{P}_{\rm em}\sim 1$~W, $\mathcal P_{\rm
detectable}=10^{-20}$~W, $\nu_0=1.3\ {\rm GHz}$) in the text. In
both cases we have used $|G|=1$ for $\mg\leq\omega_{0}$ and $|G|=0$
for $\mg>\omega_{0}$, for simplicity,  for the ``geometry
factor''~\eqref{geofac}. The thin dashed dotted line corresponds to
the sensitivity which one might get from the optimistic scenario, if
one scans the frequency between $250\ {\rm MHz}\lesssim
\nu_0\lesssim 250\ {\rm GHz}$, corresponding to $1\,\mu {\rm eV}
\lesssim \omega_0\lesssim   1\,{\rm meV}$ (for frequencies
$\nu_{0}>3\,{\rm{GHz}}$, the losses in the cavities grow due to an
increased surface resistance \cite{Aune:2000gb}; accordingly, we
have assumed a drop in the $Q$ value for frequencies higher than $3
\,{\rm{GHz}}$.) \hfill \label{limits}}
\end{center}
\end{figure}

Let us now switch to a field basis in which the prediction of $\gamma$--$\gamma^\prime$ oscillations
becomes apparent.
In fact, the kinetic terms in the Lagrangian~(\ref{lagrangian})
can be diagonalized by a shift
\begin{equation}
\label{shift}
B^{\mu}\rightarrow \tilde{B}^{\mu}-\chi A^{\mu}.
\end{equation}
Apart from a multiplicative renormalization of the electromagnetic gauge coupling,
$e^2\rightarrow e^2/(1-\chi^2)$, the visible-sector fields remain
unaffected by this shift and one obtains a non-diagonal mass term that mixes photons
with paraphotons,
\begin{equation}
\label{masssimple}
{\mathcal{L}}=-\frac{1}{4} F^{\mu\nu}F_{\mu\nu}-\frac{1}{4}\tilde{B}^{\mu\nu}\tilde{B}_{\mu\nu}
+\frac{1}{2}m_{\gamma^\prime}^2 \left(\tilde{B}^{\mu}\tilde{B}_{\mu}-2\chi \tilde{B}^{\mu}A_{\mu}+
\chi^2 A^{\mu}A_{\mu}\right) .
\end{equation}
Therefore, in analogy to neutrino flavour oscillations, photons may oscillate in vacuum into
paraphotons.
These oscillations and the fact that the paraphotons do not interact with ordinary matter
forms the basis of the possibility~\cite{Okun:1982xi} to search for signals of paraphotons in
``light shining through a wall'' experiments (cf.~Fig.~\ref{lswnomagnetic}). The sensitivity
of ongoing experiments of this type (for a review, see Ref.~\cite{Ringwald:2006rf})
for paraphoton searches has recently been estimated in Ref.~\cite{Ahlers:2007rd}. Their discovery
potential extends the current upper limit on $\chi$ set by the BFRT collaboration~\cite{Cameron:1993mr}
by about one order of magnitude over the whole range of
masses $m_{\gamma^\prime}$ (cf. Fig.~\ref{limits}).

\begin{figure}[t]
\begin{center}
\includegraphics*[width=.85\textwidth]{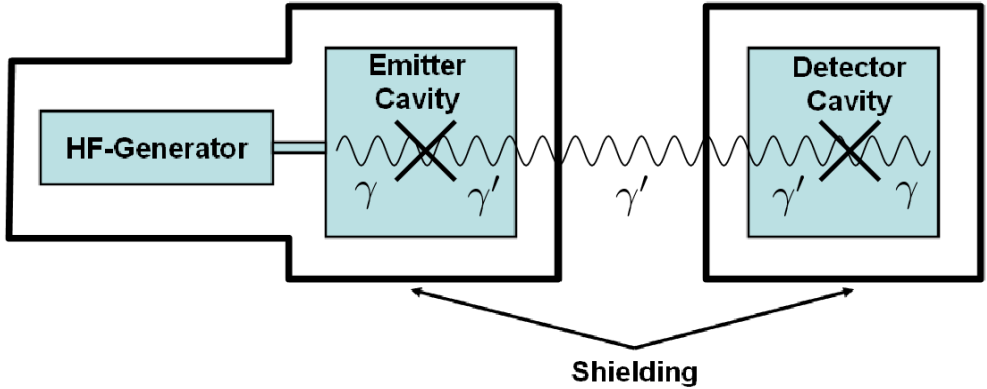}
\end{center}
\vspace{-0.5cm} \caption{\small
Schematic illustration of a
``microwaves permeating through a shielding'' experiment
 for the search for massive hidden sector photons mixing
with the photon  (a high-frequency
(HF) generator drives the emitter cavity).}
\label{cavityexp}
\end{figure}

Here, we propose another setup searching for signatures of $\gamma$--$\gamma^\prime$ oscillations
which resembles the classic ``light shining through a wall'' setup. It consists of two microwave cavities
shielded from each other (cf. Fig.~\ref{cavityexp}).
In one cavity, paraphotons are produced
via photon-paraphoton oscillations. The second, resonant, cavity is then driven by the paraphotons that
permeate the shielding and reconvert into photons. Due to the high quality factors achievable for
microwave cavities and the good sensitivity of microwave detectors such a setup will allow for
an unprecedented discovery potential for hidden sector photons in the mass range  from
$\mu{\rm{eV}}$ to ${\rm{meV}}$ range (cf.~Fig.~\ref{limits}).

Before we start with a detailed calculation, let us present a simple
estimate based on a comparison with the familiar ``light shining
through a wall'' setup, which exploits an optical cavity and a laser
with wavelength $\sim {\rm{few}} \times 100\,{\rm{nm}}$. In optical
cavities, the spatial extent of the laser beam transverse to the
beam direction ($\sim {\rm{mm}}-{\rm{cm}}$) is much greater than the
wavelength. The wave is effectively a plane wave propagating in the
beam direction and the problem is effectively one-dimensional. This
is not the case for microwave or radio-frequency (RF) cavities where
the size of the cavity is similar to the wavelength in all three
directions. Nevertheless, let us for the moment imagine an
unrealistic cavity which has infinite extent in two directions. Then
the situation is equivalent to a standard ``light shining through a
wall'' experiment (the shielding is equivalent to a wall). For a
setup with cavities on both sides of the wall, the probability for a
photon to pass through the wall and to be emitted by the second
cavity is~\cite{Okun:1982xi,Hoogeveen:1990vq,Sikivie:2007qm,Ahlers:2007rd},
\begin{equation}
\label{onedimensional}
P_\text{trans} =
16 \chi^4\left[\frac{N_{1}+1}{2}\right]\left[\frac{N_{2}+1}{2}\right]\left[\sin\left(\Delta k \ell_{1}\right)
\sin\left(\Delta k \ell_{2}\right)\right]^2.
\end{equation}
Here, $N_{1,2}$ are the number of passes the light makes through the cavities, $\ell_{1,2}$ are
the lengths of the two
cavities, and
\begin{equation}
\Delta k=\omega-\sqrt{\omega^{2}-\mg^2}
\end{equation}
is the momentum difference between the photon and the paraphoton, expressed in terms of
the energy of the laser photons, $\omega =2\pi\nu$, where $\nu$ is the frequency of the laser light.
Maximal sensitivity  to the mixing parameter $\chi$ will be achieved if both
sines in Eq.~(\ref{onedimensional}) are equal to one.
One way to achieve this is to choose  the angular frequency
 $\omega\equiv 2\pi/\lambda = \mg$ and in exploiting cavities of length
$\ell_{1}=\ell_{2} = \lambda/2$, where $\lambda$ is the wavelength of the
laser light.

Using this we can get a rough idea of what may be accomplished by a similar
experiment using microwave or RF cavities (cf. Fig.~\ref{cavityexp})
instead of optical cavities.
Using $(N+1)/2\sim Q$, where $Q$ is the quality factor of the cavity, we roughly expect
\begin{equation}
\label{onedimensionalmax}
P^{\rm{max}}_\text{trans} \sim \chi^4 Q Q^{\prime}.
\end{equation}
To get an idea of the sensitivity which such an experiment can reach let us plug in some
numbers. The power output
${\mathcal{P}}_{\rm{out}}$ of the detector cavity\footnote{When we speak of power going into and
out of the cavities we can
alternatively think of this as a measure for the energy stored inside the cavity. The power is related to
the stored energy $U$ and the $Q$ factor via
${\mathcal{P}}=\omega U/Q$.} will be
\begin{equation}
\label{output}
{\mathcal{P}}_{\rm{out}}=P_{\rm{trans}}{\mathcal{P}}_{\rm{in}},
\end{equation}
in terms of the power ${\mathcal{P}}_{\rm{in}}$ put into the emitter cavity.
An input power of ${\mathcal{P}}_{\rm{in}}\sim 1$~W
is quite realistic\footnote{Let us express this in terms of the energy stored inside the cavity.
For example at the frequency $1.3\,{\rm{GHz}}$
used in the TESLA cavities and with a $Q\sim 10^{11}$~\cite{Lilje:2004ib} this corresponds to an energy
$U\sim 10\,{\rm{J}}$ stored inside the cavity.}~\cite{Knabbe:priv} and an emission of
${\mathcal{P}}_{\rm{out}}\sim 10^{-26}$~W
is just on the verge of being detectable~\cite{Asztalos:2003px}.
High quality cavities based on superconducting technology
can reach $Q\sim 10^{11}$~\cite{Lilje:2004ib}. Plugging these numbers into
Eqs.~\eqref{onedimensionalmax} and \eqref{output}, we infer that, very optimistically, a setup
based on microwave or RF cavities might be sensitive to values of the mixing parameter as small as
$\chi\sim 10^{-12}$, in the mass range corresponding to the frequency range, i.e.
$\mu{\rm eV}\lesssim m_{\gamma^\prime}\lesssim {\rm meV}$.

Motivated by this estimate, let us proceed to more realistic situations,
taking into account the appropriate fully three-dimensional geometry.
Our starting point are the equations of motion for the
photon field $A$ and the paraphoton field $B$ (Lorentz indices
suppressed\footnote{Although we may think of $A$ as the
gauge potential.
Using Coulomb gauge $A^{0}=0$, which is compatible with Lorentz gauge, we can immediately relate
$A$ to electric fields
via $E=-\frac{dA}{dt}$.})
following from Eq.~\eqref{lagrangian},
\begin{eqnarray}
\label{photon}
(\partial^{\mu}\partial_{\mu}+\chi^{2}\mg^{2})A\!\!&=&\!\!\chi \mg^2
B,
\\
\label{paraphoton}
(\partial^{\mu}\partial_{\mu}+\mg^{2})B\!\!&=&\!\!\chi \mg^2 A.
\end{eqnarray}
Our strategy is as follows (cf. Fig.~\ref{cavityexp}).
We start with the ordinary electromagnetic field inside the first
``emitter'' cavity.
This field acts as a source for the paraphoton field.
The paraphoton field permeates the shielding and in turn acts as a source
for an electromagnetic field inside the
second ``detector'' cavity.
We will always consider the lowest non-trivial order.

To lowest order in $\chi$,
we can obtain the electromagnetic field inside the emitter cavity by
solving $\partial^{\mu}\partial_{\mu}A=0$, i.e. the standard equation of ordinary electrodynamics.
Implementing the (time independent) boundary conditions of the cavity is a textbook
exercise~\cite{Jackson}.
Using the separation
ansatz
\begin{equation}
\label{separation}
A_{\rm{em}}(\mathbf{x},t)=a_{\rm{em}}(t) A_{\omega_{0}}(\mathbf{x}),
\end{equation}
accounting for a finite quality factor $Q$ of the cavity, and including a driving force $f(t)$, we have,
\begin{equation}
\left(\frac{d^2}{dt^2}+\frac{\omega_{0}}{Q}\frac{d}{dt}+\omega^{2}_{0}\right)a_{\rm{em}}(t)=f(t),
\end{equation}
where
\begin{equation}
-\nabla^{2}A_{\omega_{0}}(\mathbf{x})=\omega^{2}_{0}A_{\omega_{0}}(\mathbf{x})
\end{equation}
is an eigenfunction of the spatial part of the wave equation including the appropriate boundary conditions.
It is convenient to choose a normalization,
\begin{equation}
\int_{V} d^{3}\,\mathbf{x} |A_{\omega_{0}}(\mathbf{x})|^{2}=1.
\end{equation}
For example, if the cavity is a cube with side length $L$,
the eigenfunctions for the electric field in the $z$-directions are
\begin{equation}
\label{cubic}
A_{\omega^{mnp}_{0}}(\mathbf{x})=C_{mnp} \sin\left(\frac{m\pi x}{L}\right)
\sin\left(\frac{n\pi y}{L}\right)\cos\left(\frac{p\pi z}{L}\right),
\end{equation}
where $C_{mnp}$ are normalization factors. The eigenvalues are in this case
given by
\begin{equation}
\omega^{mnp}_{0}=\frac{\pi}{L} \sqrt{m^2+n^2+p^2},\quad\quad m,n=1,2,\ldots\quad\quad p=0,1,\ldots\,.
\end{equation}

Employing an oscillating driving force,
\begin{equation}
f(t)=f_{0}\exp(-{\rm i}\omega t),
\end{equation}
the amplitude will eventually approach,
\begin{equation}
\label{asymptotic}
a_{\rm{em}}(t)=a^{0}_{\rm{em}}\exp(-{\rm i}\omega t)= \frac{f_{0}}{\omega^{2}-\omega^{2}_{0}-{\rm i}
\frac{\omega \omega_{0}}{Q}}\exp(-{\rm i}\omega t).
\end{equation}
For $Q\gg 1$ and
a driving force that is resonant with the cavity,  $\omega=\omega_{0}$,
the amplitude is enhanced by a factor
of $Q$ with respect to the driving force,
\begin{equation}
a^{0}_{\rm{em}}\approx   {\rm i} \frac{Q}{\omega^{2}_{0}} f_{0}.
\end{equation}

The field $a_{\rm{em}}(t)A_{\omega_{0}}(\mathbf{x})$ now acts as a source on the right hand side
of the equation of
motion for the paraphoton fields, Eq. \eqref{paraphoton}.
The paraphoton does not interact with ordinary matter and no boundary conditions are enforced at
finite $\mathbf{x}$.
The appropriate solution are therefore
obtained from the (retarded) massive Greens function,
\begin{equation}
\label{parafield} B(\mathbf{x},t) =\chi\mg^2 \int_{V} d^{3}
\mathbf{y}\frac{\exp({\rm i} k |\mathbf{x}-\mathbf{y}|)}{4\pi
|\mathbf{x}-\mathbf{y}|}
 a_{\rm{em}}(t) A_{\omega_{0}}(\mathbf{y}),
\end{equation}
where $V$ is the volume of the emitter cavity and
\begin{equation}
k^2=\omega^2 -\mg^2.
\end{equation}

In our detector cavity,
the field $B(\mathbf{x},t)$ now acts as a source, i.e. a driving force.
The wave equation can again be solved by a separation ansatz analog to Eq. \eqref{separation},
$A^{\prime}(\mathbf{x},t)=a_{\rm{det}}(t)A^{\prime}_{\omega^{\prime}_{0}}(\mathbf{x})$,
\begin{equation}
\label{detectorcavity}
\left(\frac{d^2}{dt^2}+\frac{\omega^{\prime}_{0}}{Q^{\prime}}\frac{d}{dt}+\omega^{\prime\,2}_{0}\right)
a_{\rm{det}}(t)=b(t).
\end{equation}
The driving force $b(t)$ can be obtained by remembering that the spatial eigenfunctions of cavities
form a complete orthonormal set. Inserting the separation ansatz into Eq. \eqref{paraphoton},
multiplying by the
eigenfunction $A^{\prime}_{\omega^{\prime}_{0}}(\mathbf{x})$ and integrating over the volume $V^{\prime}$
of the
detector cavity, we find
\begin{equation}
\label{drivingforce} b(t)=\chi^2 \mg^4
a_{\rm{em}}(t)\int_{V^{\prime}}\int_{V}d^{3}\mathbf{x}\,d^{3}\mathbf{y}\,
\frac{\exp({\rm i} k
|\mathbf{x}-\mathbf{y}|)}{4\pi|\mathbf{x}-\mathbf{y}|}
A_{\omega_{0}}(\mathbf{y})A^{\prime}_{\omega^{\prime}_{0}}(\mathbf{x}).
\end{equation}
To get resonant enhancement we choose
\begin{equation}
\omega^{\prime}_{0}=\omega_{0}.
\end{equation}

\begin{figure}[t]
\begin{center}
\begin{picture}(200,140)
\Text(-7,128)[l]{\scalebox{1.3}[1.3]{$|G|$}}
\Text(195,-8)[l]{\scalebox{1.3}[1.3]{$k/\omega_{0}$}}
\includegraphics[width=0.5\linewidth]{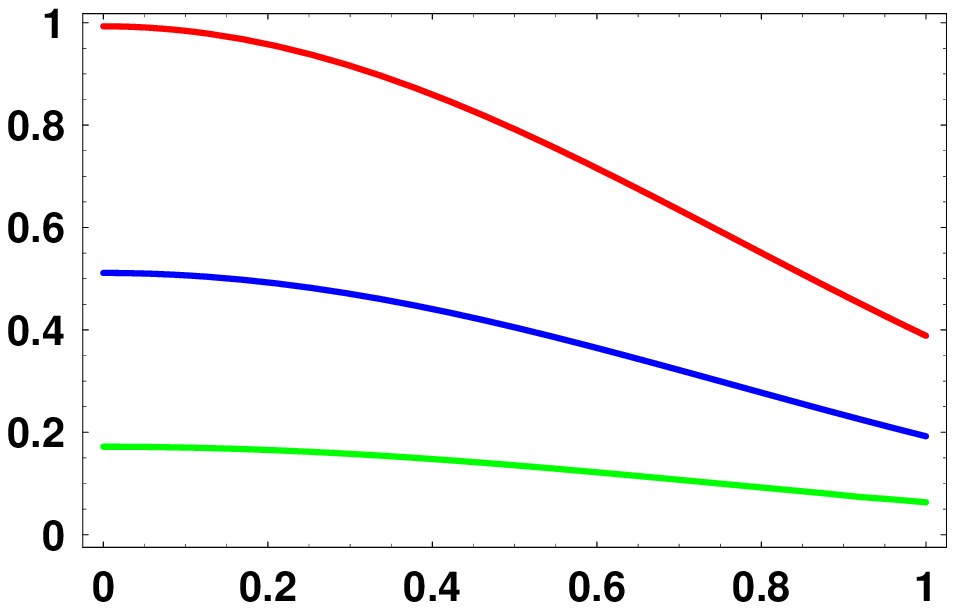}
\end{picture}
\end{center}
\caption{\small Geometry factor $|G|$ for a setup with two identical cubic cavities with side length
$L=\sqrt{2}\pi/\omega_{0}$
in the $n=1,m=1,p=0$ mode of Eq. \eqref{cubic}. The cavities are placed parallel and are separated by a distance
$d=0$ (red), $d=L$ (blue) and
$d=5\,L$ (green) along the z-axis. As expected $|G|$ scales roughly with $1/d$.}\label{formfactor}
\end{figure}

The integral in Eq. \eqref{drivingforce} has dimensions $\rm{length}^2=\rm{frequency}^{-2}$.
Taking this into account
we write
\begin{equation}
\label{detectorforce}
b(t)=a_{\rm{em}}(t)\frac{\chi^2\mg^4}{\omega^{2}_{0}}
G(k/\omega_{0})
\end{equation}
where $G$ is a dimensionless function that encodes the geometric
details of the setup, e.g. relative position, distance and shapes of
the cavities. Moreover, it depends on the mass $\mg$ via $k$,
\begin{equation}
\label{geofac}
G(k/\omega_{0})\equiv \omega^{2}_{0}\int_{V^{\prime}}\int_{V}d^{3}\mathbf{x}\,d^{3}\mathbf{y}\,
\frac{\exp({\rm i} k |\mathbf{x}-\mathbf{y}|)}{4\pi|\mathbf{x}-\mathbf{y}|}
A_{\omega_{0}}(\mathbf{y})A^{\prime}_{\omega_{0}}(\mathbf{x}).
\end{equation}
It is typically of order one,  as can be seen from Fig.~\ref{formfactor},
where we show $G$ for a setup with two identical
cubic cavities.

After some time the amplitude in the detector cavity will approach
\begin{equation}
a^{0}_{\rm{det}}= {\rm i}Q^{\prime}\frac{\chi^2\mg^4}{\omega^{4}_{0}} G \,
a^{0}_{\rm{em}}.
\end{equation}

Finally, we have to relate the amplitudes to the power input/output
in the emitter/detector cavities.
The quality factor
is directly related to the power consumption/emission of a cavity,
\begin{equation}
{\mathcal{P}}=\frac{\omega_{0}}{Q} U,
\end{equation}
where
\begin{equation}
U=const\, |a|^2
\end{equation}
is the energy stored inside the cavity.
Using this relation,
the probability for a photon to pass through the shielding and to be emitted by the second cavity is
\begin{equation}
\label{ptrans}
P_{\rm{trans}}=\frac{{\mathcal{P}}_{\rm{det}}}{{\mathcal{P}}_{\rm{em}}}
=  \frac{Q}{Q^{\prime}}\frac{|a^{0}_{\rm{det}}|^{2}}{|a^{0}_{\rm{em}}|^{2}}
= \chi^4\,  Q Q^{\prime}\, \frac{\mg^{8}}{\omega^{8}_{0}}\, |G|^2.
\end{equation}
If we choose the cavity frequency to be $\omega_{0}=\mg$, our
expression agrees up to a factor of order unity with our estimate~\eqref{onedimensionalmax}.

Let us now turn to the $\mg$ dependence of the effect.
As can be seen from Fig.~\ref{formfactor},
 the $k$ and therefore $\mg$ dependence of $|G|$ is
not very strong. Moreover, the latter is non-zero\footnote{It should be mentioned that
this is typical for the lowest cavity modes where the field does not
change sign inside the cavity.} for all allowed $\omega_{0}\geq
k\geq 0$.  Therefore, the sensitivity of the proposed experiment to the value of the mixing
parameter $\chi$
decreases roughly as $\omega_0^2/\mg^{2}$ when going to smaller masses.  What
happens for $\mg>\omega_{0}$? In this case $k={\rm i}\kappa$ is imaginary
and $|G|$ drops exponentially as $|G|\sim \exp(-\kappa d)$ where $d$
is the distance between the cavities. Since some distance between
the cavities is necessary to allow for shielding etc., the experiment
will not be very sensitive in this region.

In Fig.~\ref{limits}, we
sketch the sensitivity region of two scenarios:
\begin{itemize}
\item{} An optimistic scenario where we basically stick together the best cavities
$Q=Q^{\prime}\sim 10^{11}$, $\mathcal{P}_{\rm em}\sim 1$~W,
best detectors ${\mathcal{P}}_{\rm{detectable}}\sim 10^{-26}$~W, and assume
that perfect shielding is possible and
\item{} a more modest
scenario where we use $Q\sim 10^{10}$, $Q^{\prime}\sim 10^4$, $\mathcal{P}_{\rm em}\sim 1$~W,
 and a detectable power of
${\mathcal{P}}_{\rm{detectable}}\sim 10^{-20}$~W.
\end{itemize}
In both cases the achieved sensitivity is better than the current laboratory \emph{and} astrophysical limits.
The sensitivity region can be further extended by performing several experiments at different frequencies or,
even better, scanning through a whole region
of frequencies (thin dashed dotted line in Fig.~\ref{limits}).

At last we note that one can also obtain a bound from the observed
maximal $Q$ value of the emitter cavity itself. The conversion of photons into
paraphotons leads to an energy loss in the emitter cavity. From our discussion above we can estimate that
the probability for a photon to convert to a paraphoton during one pass through the cavity is
\begin{equation}
P_{\rm{loss}}\sim  \chi^2 \frac{\mg^4}{\omega^{4}_{0}}.
\end{equation}
Assuming that conversion into paraphotons is the only source of energy loss we infer that the maximal
possible $Q$ is
\begin{equation}
Q^{\rm{max}}\sim \frac{2\pi}{P_{\rm{loss}}}\sim \frac{2\pi}{\chi^2}\frac{\omega^{4}_{0}}{\mg^{4}}.
\end{equation}
At the resonance frequency, $\omega_{0}=\mg$, an observed value
$Q>10^{11}$  will
then translate into a bound of roughly $\chi(\mg = \omega_0 )\lesssim 10^{-5}$.
 As above, the bound becomes weaker as
$\sim \omega^{2}_{0}/\mg^2$,
for smaller $\mg$, and drops off sharply for $\mg>\omega_{0}$.

Finally, let us comment on a few experimental issues. First, since
we use cavities both for production and detection we have to assure
that both cavities have the same resonant frequency. More precisely
the frequencies have to agree in a small range
$\Delta\omega_{0}/\omega_{0}\sim 1/Q^{\prime}$. This is a
non-trivial task. However, compared to optical frequencies (as
proposed in \cite{Hoogeveen:1990vq,Sikivie:2007qm}), this should be
significantly simpler for microwave of RF cavities: the wavelength
is longer and correspondingly the allowed inaccuracies in the cavity
are much larger. Indeed, the cavities originally developed for the
TESLA accelerator~\cite{Lilje:2004ib} may be mutually tuned in
frequencies to a ${\rm{few}}\times 100\,\rm{Hz}$~\cite{Knabbe:priv}.
With a resonance frequency of roughly 1 GHz, this corresponds to an
allowed quality factor of the detector cavity of $Q^{\prime}\sim
10^{6}$. We have used even a somewhat smaller $Q^{\prime}\sim 10^4$
in our modest scenario. Second, one needs to provide sufficiently
good shielding between the cavities to prevent exciting the detector
cavity by ordinary electromagnetic fields leaking from the
production region. This is closely linked to the question how one
can decide that a possible signal is physical in origin. One way to
accomplish this  could consist in checking the phase of a
``signal''. The phase differs between an artifact resulting from a
ordinary photon sneaking out of the cavity and a true paraphoton
signal: for a true signal the phase is encoded in the complex phase
of $G$. The photon is massless and the wavenumber is
$k_{\gamma}=\omega_{0}$ whereas the paraphoton is massive and has a
smaller wavenumber
$k=\sqrt{\omega^{2}_{0}-\mg^2}<\omega_{0}=k_{\gamma}$. Therefore,
the phase difference, $\Delta \phi$, between an artifact and a true
signal is approximately
\begin{equation}
\Delta \phi \sim (k_{\gamma}-k)d=(\omega_{0}-k)d,
\end{equation}
where $d$ is the distance between the
cavities\footnote{One might also wonder how one can distinguish this signal
from a signal caused by an electric current of minicharged
particles, the latter being Schwinger pair produced in the
electric field of the cavity,
 as suggested in Ref.~\cite{Gies:2006hv}.
This is actually quite simple. Since such a current would
flow in the direction of the electric field one can simply choose the separation between the
two cavities in our
setup to be perpendicular to
the electric field. The minicharged particle current would then simply miss the detector cavity.}.

\begin{figure}[t]
\begin{center}
\includegraphics*[width=.85\textwidth]{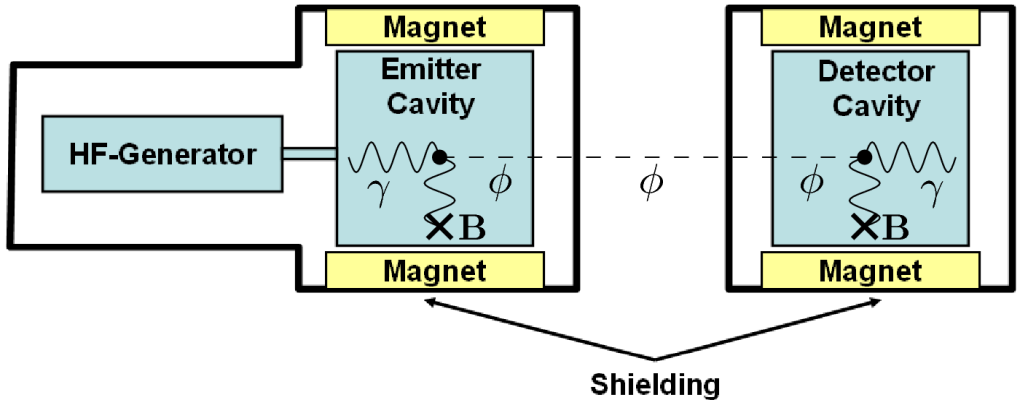}
\end{center}
\vspace{-0.5cm} \caption{\small
Schematic illustration of a
``microwaves permeating through a shielding'' experiment
 for the search for an axion-like particle $\phi$ mixing
with the photon in the presence of a magnetic field.}
\label{cavityalpexp}
\end{figure}

Last, but not least, let us note that the experimental setup proposed in this letter can
be extended~\cite{Hoogeveen:1992uk} to a search facility for light neutral spin-zero (axion-like)
particles $\phi$, coupling to electromagnetism, at low energies, according to
\begin{equation}
\label{em_pseudoscalar}
{\mathcal L} = -\frac{1}{4} F_{\mu\nu}F^{\mu\nu} + \frac{1}{2} \partial_\mu\phi\partial^\mu\phi
-\frac{1}{2}m_\phi^2\phi^2 -\frac{1}{4} g\phi F_{\mu\nu}{\tilde F}^{\mu\nu}
\,,
\end{equation}
where ${\tilde F}_{\mu\nu}$ is the dual electromagnetic field strength
tensor\footnote{The effective Lagrangian~(\ref{em_pseudoscalar})
applies for a pseudoscalar $\phi$, corresponding to a parity odd spin-zero boson. In the case
of a scalar axion-like particle, the $F_{\mu\nu}{\tilde F}^{\mu\nu}$ in Eq.~(\ref{em_pseudoscalar}) is replaced
by  $F_{\mu\nu}{F}^{\mu\nu}$.}. In fact, by placing both the emitter as well as the detector
cavity each into a magnet of strength $B$, one may drive the detector cavity
now with the axion-like particles which have been produced in the emitter cavity and which
have reconverted in the detector cavity  (cf. Fig.~\ref{cavityalpexp}). From a calculation
similar to the one presented in this letter, one finds, in analogy to \eqref{ptrans}, for
the probability that a photon passes through the shielding and is emitted by the
second cavity~\cite{Hoogeveen:1992uk},
\begin{equation}
\label{ptransalp}
P_{\rm{trans}}
\sim \left( \frac{g\, B}{\omega_0}\right)^4\,  Q Q^{\prime}\, |G|^2.
\end{equation}
Let us estimate the discovery potential of such an experiment, given
current technology. State-of-the-art axion dark matter experiments
such as ADMX~\cite{Asztalos:2003px} exploit standalone RF cavities
based on normal-conducting technology\footnote{In the strong
magnetic field required for this setup it is unclear whether one can
use a superconducting cavity that would, in principle, allow for a
higher $Q$ value.} with $Q\sim 10^6$ inside a strong magnet with
$B\sim 10$~T. Using these numbers and $P_{\rm em}=1$~W, $P_{\rm
det}=10^{-26}$~W, we estimate a sensitivity of $g\sim 8\times
10^{-10}$~GeV$^{-1}$, for $m_\phi\lesssim
\omega_0=5$~$\mu$eV -- about one order of magnitude
above the (albeit model-dependent, cf.,
e.g.,~\cite{Jaeckel:2006id,Jaeckel:2006xm}) limits  set by solar
energy loss considerations~\cite{Raffelt:1996} and by the
non-observation of solar axion-like particle induced photon
regeneration by the CAST
collaboration~\cite{Andriamonje:2007ew}, but considerably better
than the limits of present day optical ``light shining
through a wall'' experiments (for a review and references, see,
e.g., Ref.~\cite{Ringwald:2006rf}). An improvement in
the sensitivity to the range $g\sim (10^{-15}$ to
$10^{-14})$~GeV$^{-1}$, predicted for proper QCD
axions~\cite{Weinberg:1978ma,Wilczek:1977pj,Peccei:1977hh} in the
$\mu$eV mass
range~\cite{Bardeen:1977bd,Kaplan:1985dv,Srednicki:1985xd}, will
require still substantial technical advances.

\emph{In conclusion:} In this letter,
we have proposed a simple experiment to search for massive
hidden sector
photons that have kinetic mixing
with ordinary photons. The experiment would allow to probe a region of parameter space that is so
far unexplored
by laboratory experiments \emph{as well as}
astrophysical observations. Therefore, it
bears significant discovery potential for hidden sector physics.

\section*{Acknowledgment}

We would like to thank Holger Gies, Ernst-Axel Knabbe and Lutz Lilje for many discussions
and important information.


\begin{thebibliography}{10}



\bibitem{Yao:2006px}
  W.~M.~Yao {\it et al.}  [Particle Data Group],
  J.\ Phys.\ G {\bf 33} (2006) 1.



\bibitem{Okun:1982xi}
  L.~B.~Okun,
  Sov.\ Phys.\ JETP {\bf 56} (1982) 502
  [Zh.\ Eksp.\ Teor.\ Fiz.\  {\bf 83} (1982) 892].


\bibitem{Holdom:1985ag}
  B.~Holdom,
  Phys.\ Lett.\  B {\bf 166} (1986) 196.



\bibitem{Williams:1971ms}
  E.~R.~Williams, J.~E.~Faller and H.~A.~Hill,
  Phys.\ Rev.\ Lett.\  {\bf 26} (1971) 721.

\bibitem{Bartlett:1988yy}
  D.~F.~Bartlett and S.~Loegl,
  Phys.\ Rev.\ Lett.\  {\bf 61} (1988) 2285.

\bibitem{Cameron:1993mr}
  R.~Cameron {\it et al.} [BFRT Collaboration],
  Phys.\ Rev.\  D {\bf 47} (1993) 3707.

\bibitem{Ahlers:2007rd}
  M.~Ahlers, H.~Gies, J.~Jaeckel, J.~Redondo and A.~Ringwald,
  arXiv:0706.2836 [hep-ph].

\bibitem{Popov:1991}
  V.~V.~Popov and O.~V.~Vasil'ev,
  Europhys.\ Lett.\  {\bf 15} (1991) 7.

\bibitem{Popov:1999}
  V.~Popov,
  Turk.\ J.\ Phys.\  {\bf 23} (1999) 943.

\bibitem{Quevedo:2002fc}
  F.~Quevedo,
  ``Phenomenological aspects of D-branes,''
{\it ICTP Spring School on Superstrings and Related Matters, Trieste, Italy, 18-26 Mar 2002,}
 published in {\it Trieste 2002, Superstrings and related matters,} 232-330pp.

\bibitem{Abel:2004rp}
  S.~Abel and J.~Santiago,
  J.\ Phys.\ G {\bf 30} (2004) R83
  [hep-ph/0404237].

\bibitem{Blumenhagen:2006ci}
  R.~Blumenhagen, B.~K\"ors, D.~L\"ust and S.~Stieberger,
  hep-th/0610327.

\bibitem{Marchesano:2007de}
  F.~Marchesano,
  hep-th/0702094.

\bibitem{Buchmuller:2006ik}
  W.~Buchm\"uller, K.~Hamaguchi, O.~Lebedev and M.~Ratz,
  hep-th/0606187.

\bibitem{Kim:2007mt}
  J.~E.~Kim, J.~H.~Kim and B.~Kyae,
  hep-ph/0702278.

\bibitem{Blumenhagen:2000wh}
  R.~Blumenhagen, L.~G\"orlich, B.~K\"ors and D.~L\"ust,
  JHEP {\bf 0010} (2000) 006
  [hep-th/0007024].

\bibitem{Aldazabal:2000sa}
  G.~Aldazabal, L.~E.~Ibanez, F.~Quevedo and A.~M.~Uranga,
  JHEP {\bf 0008} (2000) 002
  [hep-th/0005067].



\bibitem{Dienes:1996zr}
  K.~R.~Dienes, C.~F.~Kolda and J.~March-Russell,
  Nucl.\ Phys.\  B {\bf 492} (1997) 104
  [hep-ph/9610479].

\bibitem{Lust:2003ky}
  D.~L\"ust and S.~Stieberger,
  hep-th/0302221.

\bibitem{Abel:2003ue}
  S.~A.~Abel and B.~W.~Schofield,
  Nucl.\ Phys.\  B {\bf 685} (2004) 150
  [hep-th/0311051].


\bibitem{Batell:2005wa}
  B.~Batell and T.~Gherghetta,
  Phys.\ Rev.\  D {\bf 73} (2006) 045016
  [hep-ph/0512356].

\bibitem{Blumenhagen:2006ux}
  R.~Blumenhagen, S.~Moster and T.~Weigand,
  Nucl.\ Phys.\  B {\bf 751} (2006) 186
  [hep-th/0603015].

\bibitem{Abel:2006qt}
  S.~A.~Abel, J.~Jaeckel, V.~V.~Khoze and A.~Ringwald,
  hep-ph/0608248.

\bibitem{Stueckelberg:1938}
E.~C.~G.~St\"uckelberg,
Helv.\ Phys.\ Acta {\bf 11} (1938) 225.



\bibitem{Ringwald:2006rf}
  A.~Ringwald,
  hep-ph/0612127.

\bibitem{Aune:2000gb}
  B.~Aune {\it et al.},
  Phys.\ Rev.\ ST Accel.\ Beams {\bf 3} (2000) 092001
  [physics/0003011].

\bibitem{Hoogeveen:1990vq}
  F.~Hoogeveen and T.~Ziegenhagen,
  Nucl.\ Phys.\  B {\bf 358} (1991) 3.

\bibitem{Sikivie:2007qm}
  P.~Sikivie, D.~B.~Tanner and K.~van Bibber,
  Phys.\ Rev.\ Lett.\  {\bf 98} (2007) 172002
  [hep-ph/0701198].

\bibitem{Lilje:2004ib}
  L.~Lilje {\it et al.},
  Nucl.\ Instrum.\ Meth.\  A {\bf 524} (2004) 1
  [physics/0401141].

\bibitem{Knabbe:priv}
E.-A.~Knabbe, private communications.

\bibitem{Asztalos:2003px}
  S.~J.~Asztalos {\it et al.} [ADMX Collaboration],
  Phys.\ Rev.\  D {\bf 69} (2004) 011101
  [astro-ph/0310042].


\bibitem{Jackson}
J.~D.~Jackson, ``Classical Electrodynamics,'' John Wiley \& Sons, Inc.

\bibitem{Gies:2006hv}
  H.~Gies, J.~Jaeckel and A.~Ringwald,
  Europhys.\ Lett.\  {\bf 76} (2006) 794
  [hep-ph/0608238].

\bibitem{Hoogeveen:1992uk}
  F.~Hoogeveen,
  Phys.\ Lett.\  B {\bf 288} (1992) 195.

\bibitem{Jaeckel:2006id}
  J.~Jaeckel, E.~Masso, J.~Redondo, A.~Ringwald and F.~Takahashi,
  hep-ph/0605313.

\bibitem{Jaeckel:2006xm}
  J.~Jaeckel, E.~Masso, J.~Redondo, A.~Ringwald and F.~Takahashi,
  Phys.\ Rev.\  D {\bf 75} (2007) 013004
  [hep-ph/0610203].

\bibitem{Raffelt:1996}
G.~G.~Raffelt,
Stars As Laboratories For Fundamental Physics:
The Astrophysics of Neutrinos, Axions, and other Weakly Interacting Particles,
University of Chicago Press, Chicago, 1996.

\bibitem{Andriamonje:2007ew}
  S.~Andriamonje {\it et al.}  [CAST Collaboration],
  JCAP {\bf 0704} (2007) 010
  [hep-ex/0702006].

\bibitem{Weinberg:1978ma}
S.~Weinberg,
Phys.\ Rev.\ Lett.\  {\bf 40} (1978) 223.

\bibitem{Wilczek:1977pj}
  F.~Wilczek,
  Phys.\ Rev.\ Lett.\  {\bf 40} (1978) 279.

\bibitem{Peccei:1977hh}
  R.~D.~Peccei and H.~R.~Quinn,
  Phys.\ Rev.\ Lett.\  {\bf 38} (1977) 1440.

\bibitem{Bardeen:1977bd}
  W.~A.~Bardeen and S.~H.~Tye,
  Phys.\ Lett.\  B {\bf 74} (1978) 229.

\bibitem{Kaplan:1985dv}
  D.~B.~Kaplan,
  Nucl.\ Phys.\  B {\bf 260} (1985) 215.

\bibitem{Srednicki:1985xd}
  M.~Srednicki,
  Nucl.\ Phys.\  B {\bf 260} (1985) 689.

\end{thebibliography}
\end{document}